\documentclass[reprint,aps,prl]{revtex4-2}
\usepackage{amsmath,amssymb,amsfonts,dsfont}
\usepackage{graphicx}
\usepackage{color}
\usepackage{hyperref}
\usepackage{subfigure}
\usepackage{dcolumn}% Align table columns on decimal point
\usepackage{bm}% bold math
\usepackage{float}
\usepackage{mathtools}
\usepackage{amsthm}
\usepackage{bbm}
\usepackage{pstricks}
\usepackage{verbatim}
\usepackage[normalem]{ulem}

\usepackage[compat=1.1.0]{tikz-feynman}

\definecolor{orange}{cmyk}{0,0.5,1,0}
\definecolor{rossoCP3}{cmyk}{0,.88,.77,.40}
\definecolor{graa}{rgb}{0.8,0.8,0.8}
\definecolor{blaa}{rgb}{0.2,0.2,0.6}
\hypersetup{
	colorlinks, 
	bookmarksopen, 
	bookmarksnumbered,
	citecolor=blaa, 		%color of links to bibliography
	linkcolor=rossoCP3,	%color of internal links
	urlcolor=rossoCP3,			%color of external links 
	    }
\newcommand{\beq}{\begin{eqnarray}}
\newcommand{\eeq}{\end{eqnarray}}

\newcommand{\SU}{\mathrm{SU}}

\newcommand{\SO}{\mathrm{SO}}
\newcommand{\U}{\mathrm{U}}

\begin{document}
\title{Heavy Axion from a Confining Mirror GUT}

\author{Giacomo Cacciapaglia}
\email{cacciapa@lpthe.jussieu.fr}
\affiliation{Laboratoire de Physique Théorique et Hautes Energies (LPTHE), UMR 7589, Sorbonne Universit\'e \& CNRS, 4 place Jussieu, 75252 Paris Cedex 05, France.}
\author{Csaba Cs{\'a}ki}
\email{csaki@cornell.edu}
\affiliation{Laboratory for Elementary Particle Physics, Department of Physics, Cornell University, Ithaca, NY 14853, USA}
\author{Teng Ma}
\email{mateng@ucas.ac.cn}
\affiliation{International Centre for Theoretical Physics Asia-Pacific (ICTP-AP), University of Chinese Academy
of Sciences (UCAS), 100190 Beijing, China
} 
\affiliation{Taiji Laboratory for Gravitational Wave Universe (Beijing/Hangzhou), University of Chinese Academy of Sciences, Beijing 100049}

%%%%%%%%%%%%%%%%%%%%%%%%%%%%%%%%%%%%%%%%%%%%%%%%%%%%%%%%%%%%%%%%%%%%%

\begin{abstract}
We propose a new framework for solving the strong CP problem via a heavy axion, using mirror symmetry and grand unification. The mirror GUT sector remains unbroken and dynamically generates a calculable heavy mass scale via confinement without fine tuning. Models in this class feature a heavy axion, whose potential is less sensitive to Planck scale corrections, as well as a rich hidden sector from the confined mirror GUT. The solution to the strong CP problem remains unspoiled by the presence of additional phases in the GUT Yukawas, yet allowing the possibility of electric dipole moments within the reach of future experiments. Our proposal offers new directions in GUT model building, axion phenomenology, dark matter and cosmology.
\end{abstract}

\maketitle

The Standard Model (SM) of particle physics offers an exceptionally good description of particle interactions up to the TeV scale. Yet the strong sector (described by Quantum Chromodynamics - QCD) contains an unobserved source of CP violation, associated to the misalignment between the topological structure of the vacuum ($\theta$-term)~\cite{tHooft:1976snw,Jackiw:1976pf,Callan:1976je} and the overall phase of the quark Yukawa couplings. To pass current experimental bounds~\cite{Pospelov:2005pr,Abel:2020pzs}, the misalignment angle $\bar\theta$ should not exceed $10^{-10}$, hence implying a severe fine tuning between independent sectors of the SM.  

The most popular solution proposed by Peccei and Quinn~\cite{Peccei:1977hh,Peccei:1977ur} introduces an anomalous  $\U(1)_\text{PQ}$ global symmetry that allows one to shift away the dangerous phase. Its spontaneous breaking produces a light pseudoscalar, the axion. A potential for the axion is  generated by QCD, ensuring that the axion  dynamically relaxes to $\bar\theta=0$. The QCD axion generally has a mass of the order $m_a \sim m_\pi f_\pi/f_a$, where $f_a$ is the axion decay constant, hence being tightly constrained by astrophysics~\cite{Carenza:2024ehj,Caputo:2024oqc} to be very light, $m_a \lesssim 10^{-3}$~eV with $f_a \gtrsim 10^{9}$~GeV~\cite{AxionLimits}. Due to the light mass of the axion, the solution is extremely sensitive to generic contributions from gravity at the Planck scale, leading to  the ``axion quality'' problem~\cite{Kamionkowski:1992mf,Holman:1992us}.  The QCD axion is also an excellent candidate for ultralight dark matter (see for example \cite{Adams:2022pbo} for a recent review).

A mitigation of the above shortcomings would occur if the axion received additional contributions to its potential from scales higher than that of QCD. Several attempts have been made, e.g. Refs~\cite{Holdom:1985vx,Gaillard:2018xgk,DiLuzio:2020wdo,Hook:2019qoh,Gherghetta:2020ofz,Murayama:2026ioh}, revealing how hard it is to sufficiently lift the axion mass ($m_a \gtrsim 1$~GeV would be enough to evade astrophysical bounds~\cite{AxionLimits}) while preserving the solution to the strong CP problem. In this letter, we propose a novel approach that combines a mirror-symmetric copy of the SM embedded within a grand unification theory (GUT) together with the dynamical generation of a high scale via confinement of the mirror GUT sector. We will show how the axion mass can be lifted above the GeV scale without introducing any new fine tuning and within a predictive framework.  

\vspace{0.5cm}

The idea of using a mirror symmetry within GUTs dates back to Rubakov~\cite{Rubakov:1997vp}. In his original proposal, the mirror $\mathbb{Z}_2$ symmetry was assumed to be explicitly violated by supersymmetry breaking effects, leading to the breaking of the two GUT gauge symmetries at different scales and dynamically generating a large confinement scale in the mirror QCD sector. Constraints on this model have been studied in \cite{Fukuda:2015ana}.
While raising the axion mass becomes rather straightforward, this class of models suffers from tuning issues common to the most general class of mirror symmetric models~\cite{Albaid:2015axa}: generating different GUT and/or electroweak symmetry breaking scales using the $\mathbb{Z}_2$ breaking requires tuning; if the theory is supersymmetric,   supersymmetry breaking generally spoils the solution to the CP problem by introducing additional phases. Mirror symmetry applied directly to the SM, or parts of it, have also been considered~\cite{Berezhiani:2000gh,Hook:2019qoh,Dunsky:2023ucb,Fukuda:2026bdn}, with similar shortcomings.

\vspace{0.5cm}

Here we propose a new class of models that generate a new dynamical mass scale in the mirror sector, without any tuning or any explicit breaking of the mirror $\mathbb{Z}_2$ symmetry. The key novelty is that we do not require the mirror GUT symmetry to be broken. Instead, the new scale enhancing the axion mass is dynamically generated via the confinement of the unbroken gauge symmetry in the mirror sector. As such, it is calculable once the details of the GUT model are specified, and it naturally falls many orders of magnitudes above the QCD scale due to the faster running of the GUT coupling compared to the QCD one. The $\mathbb{Z}_2$ mirror symmetry is spontaneously broken at high scales by the GUT breaking in the visible sector, without the need for any explicit breaking. 
This spontaneous breaking of the mirror $\mathbb{Z}_2$ is achieved via a coupling between the field whose vacuum expectation value (VEV) breaks the GUT symmetry, $\Phi$, and its mirror partner $\hat \Phi$, similarly to the model in~\cite{Dunsky:2023ucb}. Hence, we introduce a mirror-symmetric potential in the form:
\begin{equation} \label{eq:LPhi}
    \mathcal{L}\supset - \lambda_\Phi \left( \mathcal{V}(\Phi) + \mathcal{V} (\hat\Phi)\right) + \xi\
    \text{Tr}[\Phi^\dagger \Phi]\ \text{Tr} [\hat\Phi^\dagger \hat\Phi]\,,
\end{equation}
where $\mathcal{V}(X) = (\text{Tr}[X^\dagger X] - \mu^2)^2$. 
If $\xi > 2 \lambda_\Phi$, this potential has two degenerate minima where either $\Phi$ or $\hat\Phi$ have a non-vanishing VEV while the other is at zero. In either minimum, one GUT sector is broken at the scale $\Lambda_\text{GUT} = \mu$, while the other one remains unbroken and its scalar field acquires a large mass. We can identify the sector with the broken GUT symmetry to be the visible sector, while the one with the unbroken GUT symmetry to be the (hidden) mirror sector. 
As long as the unbroken GUT gauge group together with its chiral fermion content is asymptotic free, it will run to strong coupling and possibly confine, thereby  generating a new (heavy) scale as well as contributions to the axion potential, naturally aligned to that of ordinary QCD. In minimal GUT models based on $\SU(5)$ and $\SO(10)$, the confinement scale will typically lie well above the QCD scale, since they have a beta function coefficient that is larger than that of QCD. Thus we naturally obtain a confinement scale that is several orders of magnitude above the GeV scale. Since the axion mass typically scales quadratically with the confinement scale, this will result in a large contribution to the axion mass, without spoiling the alignment needed to cancel the effective $\bar\theta$ angle. 

\vspace{0.5cm}

To illustrate the main features of this class of models, let us focus on a minimal non-supersymmetric $\SU(5)$ GUT scenario~\cite{Georgi:1974sy}, with its mirror sector and a Peccei-Quinn (PQ) symmetry of the KSVZ-type~\cite{Kim:1979if,Shifman:1979if}. In this class of axion models, the PQ anomaly is generated due to the presence of additional fermions, which are vector-like with respect to the GUT gauge symmetry and obtain a mass via the PQ symmetry breaking. We also want to ensure that there is a single $\U(1)_\text{PQ}$ symmetry, so that we end up with only a single axion. This PQ symmetry should however be anomalous under both the visible and the mirror gauge symmetries. We achieve this by adding a set of gauge singlet fermions ($\psi_1$) and one scalar ($\phi_1$), which connect the PQ symmetries of the two sectors, and where the latter is responsible for the PQ breaking. All fields charged under $\SU(5)$, instead, come in mirror pairs.  Since the anomaly is related to the presence of the additional fermions, the PQ charges of the SM-like fermions should be proportional to the only non-anomalous global U(1) symmetry (the same that is sometimes used for a ``flipped SU(5)" construction~\cite{Barr:1981qv,Derendinger:1983aj}). 
The field content is summarized in Table~\ref{tab:SU5model}, where the hatted fields are charged under the mirror $\SU(5)$ gauge symmetry. The PQ fermions are chosen to be in the symmetric $\bf 15 + \overline{15}$ of $\SU(5)$ so that all allowed Yukawa couplings preserve the PQ symmetry, provided the scalar field $\Phi$ responsible  for breaking the visible $\SU(5)$ to the SM is in the $\bf 75$~\cite{Hubsch:1984pg,Hubsch:1985em}, instead of the more canonical choice of the adjoint. The choice of the $\bf 75$ is motivated by the fact that no renormalizable Yukawa couplings involving it are allowed by gauge invariance, hence this scalar is not required to carry a PQ charge. We remark that the only other choice for the PQ fermions would have been $\bf 5+\overline{5}$, as all other vector-like representations of $\SU(5)$ allow a Yukawa coupling to the $\bf 75$: we focus here on the symmetric representation for simplicity of presentation.  The ``$\times$'' in the table means that those singlet fields are not doubled - they are the ones connecting the visible and the mirror sectors, and also ensuring there is a single PQ symmetry. 
The PQ anomaly, carried by the PQ fermions, is characterized by the integer
\begin{equation} \label{eq:N}
    N = \sum_\psi t_r\ \frac{q_\psi}{|q_{\phi_1}|} = 7\,,
\end{equation}
where $q_\psi$ is the PQ charge of each field and $t_r$ its integer-valued Dynkin index under $\SU(5)$, while $q_{\phi_1}$ is the PQ charge of the scalar generating the axion.

Gauge invariance allows the following Yukawa couplings:
\begin{multline}
    - \mathcal{L}_\text{Yuk} = \Big( y_u\, \psi_{10} \psi_{10} H + y_d\, \psi_{10} \psi_{\bar 5} H^\dagger + y_1\, \psi_{\bar 5} \psi_1 H + \\
     y_{15}\, \psi_{\bar 5} \psi_{15} H^\dagger + \mbox{mirror symmetric} \Big)\ + \\
    \lambda \left( \psi_{15} \psi_{\bar{15}} + \hat\psi_{15} \hat\psi_{\bar{15}} \right) \phi_1 + \xi\, \psi_1 \psi_1 \phi_1 + \mbox{h.c.} \label{eq:LYuk}
\end{multline}
 which preserve the PQ charges and where we omitted the generation indices (masses for the $\bf 15+\overline{15}$ and singlets are forbidden by the PQ symmetry).
At the scale $\Lambda_\text{GUT}\sim 10^{16}$~GeV, one of the $\SU(5)$ gauge symmetries is broken down to the SM one via a potential of the form of Eq.~\eqref{eq:LPhi} for the real $\bf 75$ scalars $\Phi$ and $\hat \Phi$, which are neutral under the PQ symmetry. Hence, the broken $\SU(5)$ sector leads to the usual GUT dynamics and generates the SM at low energies. We identify this $\SU(5)$ with the visible sector. However, the gauge symmetry remains unbroken in the other $\SU(5)$ which we identify with the mirror sector, and denote the fields in the mirror sector by hats.
The mirror scalar $\hat \Phi$ picks up a heavy mass around $\Lambda_\text{GUT}$ and decouples. We assume that $\hat H$ could however be lighter, noting that its mass is related by the mirror symmetry to the Higgs and triplet masses in the SM sector, where the $\bf 75$ allows one to implement the doublet-triplet splitting via the `missing multiplet' mechanism~\cite{Hubsch:1985em}. Below the GUT scale, the mirror $\SU(5)$ runs toward strong coupling at lower energies and eventually confines.

\begin{table}[tb!]
    \begin{tabular}{|c|c|c|c|c|c|}
\hline
 & $\SU(5)$ & $\U(1)_\text{PQ}$ & $\SU(3)_{10}$ & $\SU(3)_{5}$ & Mirror \\ \hline\hline
 $\psi_{10}^i$ & $\bf 10$ & $1$ & $3$ & $-$ & $\hat\psi_{10}$ \\
 $\psi_{\overline 5}^j$ & $\bf \bar 5$ & $-3$ & $-$ & $\bar 3$ & $\hat\psi_{\overline{5}}$ \\
 $\psi_{1}^k$ & $\bf 1$ & $5$ & $-$ & $-$ & $\times$ \\
 $\psi_{15}$ & $\bf 15$ & $1$ & $-$ & $-$ & $\hat \psi_{15}$ \\ 
 $\psi_{\overline{15}}$ & $\bf \overline{15}$ & $9$ & $-$ & $-$ & $\hat \psi_{\overline{15}}$ \\\hline
 $\Phi$ & $\bf 75$ & $0$ & $-$ & $-$ & $\hat \Phi$ \\
 $H$ & $\bf 5$ & $-2$ & $-$ & $-$ & $\hat H$ \\
 $\phi_1$ & $\bf 1$ & $-10$ & $-$ & $-$ & $\times$ \\\hline
    \end{tabular}
    \caption{\label{tab:SU5model} Matter content of the minimal mirror $\SU(5)$ GUT model with Peccei-Quinn symmetry. All fields charged under $\SU(5)$ have a mirror partner, while singlets do not (indicated by an $\times$ in the last column). The indices $i,j,k = 1,2,3$ label the SM generations. The PQ charges are the same for the mirror partners, while the generation symmetry $\SU(3)_{10} \times \SU(3)_{5}$ is sector-specific.}
\end{table}

While the PQ symmetry remains intact at the GUT scale, it will be broken by the VEV of the singlet $\phi_1$, giving rise to the axion field $a$
\begin{equation}
    \phi_1 = \frac{1}{\sqrt{2}} (v_1 + S) \ e^{i a/v_1} \,,
\end{equation}
where $v_1 = N f_a = 7 f_a$ defines the axion decay constant, as usual.
The radial mode $S$ acquires a mass around the scale $v_1$. 
At the PQ scale, masses are generated for the PQ fermions $\bf 15 + \overline{15}$ of both sectors, as well as for the singlets $\psi_1^k$:
\begin{equation}
    M_{15} = \hat M_{15} = \frac{\lambda N f_a}{\sqrt{2}}\,, \quad M_1^{ij} = \frac{\xi^{ij} N f_a}{\sqrt{2}}\,.
\end{equation}
In the mirror sector, after integrating out the mirror PQ fermions $\hat{\psi}_{15}+\hat{\psi}_{\overline{15}}$, the mirror $\SU(5)$ continues running towards strong coupling. Once the mirror confinement scale is reached, an axion potential including an axion mass is generated. At this scale, the axion potential forces $\bar\theta=0$ in both the mirror and visible sectors, due to the fact that both the mirror $\SU(5)$ and the SM $\SU(3)$ (QCD) have the same phases and the same PQ anomaly.

\vspace{0.5cm}

\begin{figure}[tbh]
 \includegraphics[width=7.9cm]{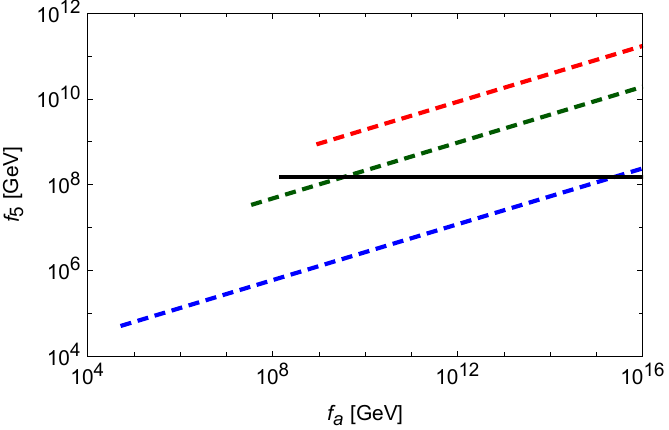}
 \caption{\label{fig1} Plot of $f_5$ as a function of $f_a$ for fixed GUT coupling in dashed lines, with $\alpha_\text{GUT} = 1/40$ (Blue), $1/30$ (Green) and $1/25$ (Red). The black solid line indicates the $f_a$-independent value $f_5^\text{run}$ obtained from the running QCD coupling. The lines ends when $f_a = f_5$.}
\end{figure}

The main advantage  of this class of models is the dynamical origin of the confinement scale $f_5$ in the mirror $\SU(5)$ sector, which can be predicted as a function of the PQ scale $f_a$ and the value of the gauge coupling at the GUT scale.
Between $\Lambda_\text{GUT}$ and $f_a$, the mirror $\SU(5)$ gauge coupling $\hat\alpha_5$ runs with the one-loop beta function:
\begin{equation}
    \frac{d \hat \alpha_5}{d\ln \mu} = - \frac{\hat{b}_5}{2\pi}\ \hat \alpha_5^2\,, \quad \mbox{with}\;\; \hat{b}_5 = \frac{29}{3}\,. 
\end{equation}
Hence, at the scale $f_a$:
\begin{equation}
    \hat \alpha (f_a) = \frac{ \alpha_\text{GUT}}{1-\frac{\hat{b}_5 \  \alpha_\text{GUT}}{2\pi} \ln \frac{\Lambda_\text{GUT}}{f_a}}\,,
    \end{equation}
where $\alpha_\text{GUT}$ is the coupling constant at the GUT scale, equal in the two sectors.
Below $f_a$, the mirror PQ fermions in the symmetric representation are integrated out and the beta function coefficient is replaced by
\begin{equation}
    \hat{b}_a = \hat{b}_5 + \frac{14}{3} = \frac{43}{3}\,.
\end{equation}
The running coupling becomes strong and triggers confinement at the scale
\begin{equation}
    f_5 = f_a\ e^{-\frac{2\pi}{\hat{b}_a \hat\alpha_5(f_a)}}\,.
\end{equation}
Note that we have neglected the contribution of $\hat H$ between $\Lambda_\text{GUT}$ and $m_{\hat H}$ for simplicity, as it remains numerically small.
Hence, given $ \alpha_\text{GUT}$, the scale of mirror $\SU(5)$ confinement $f_5$ can be predicted as a function of $f_a$, see Fig.~\ref{fig1} for some sample values with fixed $\alpha_\text{GUT}$ (dashed lines): $\alpha_\text{GUT}=1/40$ (SM running), $1/30$ and $1/25$ (supersymmetric unification).

Alternatively, assuming the SM couplings unify in the visible sector, the value of $\alpha_\text{GUT}$ itself can be estimated using the one-loop running of the QCD coupling, taking into account the effect of the PQ fermions in the symmetric representation, whose masses are of order $f_a$ in both the SM and mirror sectors. This way we can obtain an expression for $f_5$ that only depends on $\Lambda_\text{GUT}$ and the beta-function coefficient $\hat{b}_a$ (and the QCD one $b_3 = 7$) below $f_a$:
\begin{equation}
    f_5 = \Lambda_\text{GUT}\ \left( \frac{\Lambda_\text{QCD}}{\Lambda_\text{GUT}} \right)^\frac{7}{\hat{b}_a} \sim 1.54\cdot 10^8~\mbox{GeV}\,,
\end{equation} 
while $\alpha_\text{GUT}$ ranges from $0.024$ for $f_a = \Lambda_\text{GUT}$ to $0.036$ for the smallest allowed $f_a = f_5$. We will denote this fixed value as $f_5^\text{run}$, and we remark that this value does not depend on the presentation of the PQ fermions.

At the mirror confinement scale, the matter content of the mirror $\SU(5)$ sector only consists of three families of $\bf 10$ and $\bf \bar 5$ fermions.
The low energy dynamics of this model has been studied in~\cite{Csaki:2021xhi} for one family and in~\cite{Bai:2021tgl} for three chiral families (see also \cite{Li:2025tvu}, while the supersymmetric version of such a matter content was first considered in~\cite{Affleck:1983vc,Csaki:1996sm,Csaki:1996zb}). 
The non-abelian flavor symmetry is broken by some condensates, with $\SU(3)_{10}\times\SU(3)_{5} \to \SO(3)_D$ being the most likely breaking pattern. Meanwhile the global $\U(1)$ symmetry (whose charges for $\hat\psi_{10}$ and $\hat\psi_{\bar 5}$ match their respective PQ charges) remains unbroken. Hence, 't Hooft anomaly matching requires the presence of an $\SO(3)_D$ triplet of massless composite fermions $\mathcal{B}_\kappa$ generated by the composite operator $\langle \hat\psi_{10} \hat\psi_{\bar 5} \hat\psi_{\bar 5}\rangle$.  In our model, they acquire a mass via a four-fermion interaction generated by integrating out the heavy scalar $\hat H$:
\begin{equation}
    \frac{y_d^{ij}\ y_1^{lk}}{m_{\hat H}^2}\ \hat\psi_{10}^i \hat\psi_{\bar 5}^j \hat\psi_{\bar 5}^l \psi_1^k \to \frac{y_d^{ij}\ y_1^{lk}\ f_5^3}{m_{\hat H}^2}\ \zeta^{ijl}_\kappa \mathcal{B}_\kappa \psi_1^k\,,
\end{equation}
where $\zeta^{ijl}_\kappa$ is a Clebsch-Gordan coefficient and we assumed $m_{\hat H} > f_5$.
Combined with the Majorana mass for $\psi_1^i$ generated at the PQ breaking scale, $M_1^{ij}$, the model predicts six Majorana singlets, with three light ones having masses $\sim \frac{f_5^6}{m^4_{\hat H} f_a}$. Note that the singlets $\psi_1$ also couple to the left-handed neutrinos in the visible sector, however this mixing will not generate a mass for neutrinos: this feast can be straightforwardly achieved by replacing a single set of $\psi_1$ with two mirror copies, where $y_1$ must be identified with the neutrino Yukawas in a type-I see-saw mechanism. Other mechanisms can be implemented, as long as they do not spoil the PQ symmetry.

\begin{figure}[tbh]
 \includegraphics[width=7.9cm]{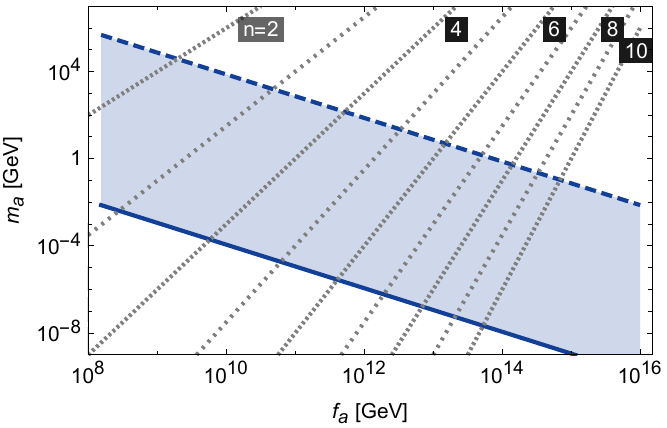}
 \caption{\label{fig2} Plot of $m_a$ as a function of $f_a$ for $\alpha_\text{GUT}$ computed via QCD running, including the effect of the $\bf 15 + \overline{15}$ with mass $f_a$.  The allowed mass range is delimited by $m_{\hat H} = \Lambda_\text{GUT}$ (solid) and $m_{\hat H} = f_5^\text{run}$ (dashed). The gray dotted lines indicate the minimal dimension $4+n$ of a Planck suppressed operator that preserves the solution without fine tuning.}
\end{figure}

From the breaking of the mirror flavor symmetry emerges a number of pseudo-Nambu-Goldstone bosons (pNGBs) that acquire mass via the Yukawa interactions. 
Integrating out $\hat H$ generates four-fermion interactions involving the $\SU(5)$-charged fermions in the mirror sector. The pNGBs from the non-abelian flavor symmetry breaking can be introduced by replacing the following fermion bilinears~\cite{Csaki:2021xhi} by a Goldstone matrix:
\begin{equation}
    \psi_{10}^i \psi_{10}^j \to B_1 f_5^3\ \Sigma_1^{ij}\,, \quad \psi_{10}^i \psi_{\bar 5}^j \to B_2 f_5^3\ \Sigma_2^{ij}\,, 
\end{equation}
where $\Sigma_1$ contains $5$ pNGBs from the breaking $\SU(3)_{10} \to \SO(3)$, while $\Sigma_2$ encodes $8$ pNGBs from $\SU(3)_{10} \times \SU(3)_5 \to \SU(3)_D$.  The coefficients $B_1$ and $B_2$ are $\mathcal{O}(1)$ form factors.
Henceforth, the terms in the effective Lagrangian responsible for the pNGB masses read:
\begin{multline} \label{eq:massop}
    - \frac{f_5^6}{m^2_{\hat H}} \left\{ \Big(B_1^2\ \left|\mbox{Tr} [y_u \Sigma_1]\right|^2 +  B_2^2\ \left|\mbox{Tr} [y_d \Sigma_2]\right|^2\Big) \right. \\
    \left. + \Big( B_1 B_2\ \mbox{Tr} [y_u \Sigma_1] \mbox{Tr} [y_d \Sigma_2] + \mbox{h.c.}\Big)\right\}\,.
\end{multline}
Including these terms in an effective low energy Lagrangian for the pNGBs and the axion, along the lines of \cite{DiVecchia:2013swa}, and considering diagonal Yukawa matrices, we can compute the axion mass:
\begin{multline}
    m_a^2 = \frac{f_5^6}{m_{\hat H}^2} \frac{1}{f_a^2} \frac{B_1 \sum_i y_{u,i} + B_2 \sum_j y_{d,j}}{B_1^{-1} \sum_i y_{u,i}^{-1} + B_2^{-1} \sum_j y_{d,j}^{-1}} \\ \sim \frac{B_1^2 f_5^6}{m_{\hat H}^2} \frac{1}{f_a^2} y_t \frac{B_2 y_u y_d}{B_1 y_u + B_2 y_d}\,.
\end{multline}
Note that the leading contribution to the axion mass is proportional to the product of the largest Yukawa coupling of the top, $y_t$, with the usual combination of up and down Yukawas, $y_u y_d/(y_u+y_d) \sim 10^{-5}$ (assuming $B_1 = B_2$). Thanks to the mirror symmetry, these couplings are equal to the SM ones at the GUT scale, and they only differ by the differential running between the SM and mirror sectors from $\Lambda_\text{GUT}$ to $f_5$. In Fig.~\ref{fig2} we show the allowed range of $m_a$ as a function of $f_a$ for the QCD running solution $f_5=f_5^\text{run} = 1.5\cdot 10^8$~GeV, varying between $m_{\hat H} = \Lambda_\text{GUT}$ (solid bottom line) and $m_{\hat H} = f_5^\text{run}$ (dashed top line). We see that axion masses above $1$~GeV can be achieved in a wide portion of the parameter space. The enhanced mass also yields a softening of the axion quality problem, which can be measured via the contribution of a generic Planck-suppressed operator of dimension $4+n$ to the misalignment of $\bar \theta$. 
A generic operator of this type is in the form 
\begin{equation}
\mathcal{O}_{4+n} = c\  \frac{(\phi_1^\dagger \phi_1)^{\frac{n+4-q}{2}} \phi_1^q}{M_{\text{Pl}}^{n}} + \text{h.c.}\,,
\end{equation}
where $c$ is a complex coupling constant of order $\mathcal{O}(1)$ that is assumed to carry an additional CP-violating phase $\delta$, uncorrelated to the other phases in the theory.
This term contributes to the axion potential:
\begin{equation}
\Delta V(a) = -\Delta \cos\left(\frac{q}{N} \frac{a}{f_a} + \delta\right),
\end{equation}
where the amplitude is $
\Delta \sim \frac{f_a^{4+n}}{M_{\text{Pl}}^{n}}.$ This extra contribution shifts the axion VEV, hence inducing a correction to the strong CP phase $\bar \theta$ of the order  
\begin{equation} 
\Delta \bar{\theta} \approx \frac{\Delta}{m_a^2 f_a^2}.
\end{equation}
Requiring this misalignment to be smaller than the experimental limit on $\bar\theta$, i.e. $\Delta\bar\theta \sim \frac{f_a^{2+n}}{m_a^2 M_\text{Pl}^n} \lesssim 10^{-10}$, allows to determine the minimal dimension $n$ that would not spoil the solution to the CP problem, as shown by the dashed gray lines in Fig.~\ref{fig2}. As a comparison, the QCD axion would need $n\geq 5$ for $f_a = 10^9$~GeV, while in our model we find a minimal $n$ between $2$ and $3$ for the same axion scale, depending on the value of $m_{\hat H}$.

The 13 pNGBs from the chiral symmetry breaking in the mirror GUT sector get a mass from Eq.~\eqref{eq:massop} proportional to $f_5^2/m_{\hat H}$, with coefficient given by a combination of Yukawa couplings. Numerically, for $B_1=B_2 = 1$ and using the values of the Yukawas at the electroweak scale we find
\begin{equation}
    m_\text{pNGB} \sim \{1.7, 
    \dots, 0.013\}~\mbox{GeV}
\end{equation}
for $f_5 = 1.54\cdot 10^{8}$~GeV and $m_{\hat H} = \Lambda_\text{GUT}$. Parametrically, the axion is lighter than the pNGBs by a factor $m_a^2/m_{pNGB}^2 \propto f_a^2/f_5^2$ , and only for 
$f_a \sim f_5$, will the lightest pNGB have a mass close to that of the axion. These light pNGBs have large couplings to the light singlet baryons, while they also couple to SM particles via operators suppressed by the the GUT or axion scales, generated by the $\Phi$ and $\phi_1$ portals, respectively. Hence, as long as the decay channel to the singlets is closed, they might give rise to a stable or long-lived Dark Matter candidate.

\vspace{0.5cm}

As we have demonstrated with our prototype $\SU(5)$ model, the axion potential is generated at a calculable scale $f_5 \gg \Lambda_\text{QCD}$. The spontaneously broken mirror symmetry ensures that the axion couples equally to QCD and the mirror $\SU(5)$, such that the $\bar\theta$ angle is canceled in both sectors, and the QCD contribution to the axion potential is naturally aligned to that of the confining mirror sector. There remain, however, some important questions to be addressed to ensure that our cancellation mechanism indeed works. The first is whether the presence of the additional Yukawa couplings in Eq.~\eqref{eq:LYuk} significantly modifies the flavor structure of the physically observable CP violating angle, $\bar\theta$. The second is about possible novel loop corrections in the SM and mirror sectors that may split the two $\bar\theta$ angles.

To answer the first question, we need to understand the spurion structure of the full set of Yukawa couplings in Eq.~\eqref{eq:LYuk}. The combination of fields that remain invariant under a spurious $\U(1)_\text{PQ}$ transformation (that includes the shifting of the $\theta$ angle) is most easily identified~\cite{Bedi:2024wqg} by considering the 't Hooft operator of an $\SU(5)$ instanton~\footnote{This does not imply that the axion potential is generated by instantons, we are  using it as a simplifying tool for finding the proper flavor invariant combinations.}:
\begin{equation}
     \mathcal{O}_\text{'t Hooft} = e^{-i\theta} (\psi_{10}^1 \psi_{10}^2 \psi_{10}^3)^3 (\psi_{\bar 5}^1 \psi_{\bar 5}^2 \psi_{\bar 5}^3) \psi_{15}^7 \psi_{\bar{15}}^7\,,
\end{equation}
where the power of each field is equal to the (integer-normalized) Dynkin index of their $\SU(5)$ representation.
In the above operator, one can then close up the fermion fields using Yukawa couplings to find the combinations of the couplings and $\theta$ that are invariant under all spurious symmetries. We find that the only possible combination is 
\begin{equation}
 e^{-i \theta}\ \text{det} y_u\ \text{det} y_d\ \lambda^7 \,, 
\end{equation}
leading to the definition of the observable and flavor-symmetry invariant $\bar\theta$ combination
\begin{equation}
    \bar \theta = \theta - \text{arg}\ \text{det}\ y_u y_d - 7\ \text{arg}\ \lambda\,.
\end{equation}
However, $\lambda$ is real, so we are left with the usual definition of $\bar{\theta}$ in the SM. Importantly, the phases of $y_1$ and $y_{15}$ do not enter here, even though they involve charged fermions. 

The second important potential issue is that the flavor and CP structure in the mirror sector (where the GUT symmetry is left unbroken) is different from that of the SM sector (where $\SU(5)$ is broken), hence loop corrections between $\Lambda_\text{GUT}$ and $f_5$ could potentially lead to misalignment between the two $\bar\theta$ parameters. Such effects could originate either from the different CP violating phases in the GUT Yukawas versus the SM ones, or from the presence of the additional Yukawas involving the singlets and PQ fermions in the $\bf 15+\overline{15}$ representations. The main distinction of the GUT Yukawa couplings $y_u$ and $y_d$
from the SM ones arises due to the fact that the quark doublets and right-handed up-type singlets are embedded in the same multiplet $\psi_{10}$. As a result the $y_u$ Yukawa matrix transforms in the symmetric representation of $\SU(3)_{10}$ (instead of being a bi-fundamental as in the SM). This implies that diagonalizing $y_u$ and $y_d$ leaves a physical CKM mixing matrix with $3$ physical phases, one of which remains physical even for the case of just two generations (unlike in the SM). Upon breaking the $\SU(5)$ in the SM sector, splitting the $\bf 10$-plet allows us to remove two such phases, hence reproducing the usual CKM matrix (the two additional phases would be carried by the couplings of the color-triplet Higgs, which we assume to have a mass at $\Lambda_\text{GUT}$ to avoid unacceptable proton decay rates). The logarithmically divergent contributions~\cite{Ellis:1978hq,Khriplovich:1993pf} to $\bar\theta$ in the two sectors can be traced via the equivalent of the Jarlskog invariants of the SM Yukawas~\cite{Jarlskog:1985ht,Wu:1985ea}. We found, using a careful spurion analysis, that the lowest dimensional CP-odd flavor invariant contains 8 SM-like Yukawa couplings: 
\begin{equation}
J_\text{GUT}=  i\ {\rm Tr} \left( y_u y_d^* y_d^T  y_u^* [y_u y_u^\dagger,y_d y_d^\dagger]\right)\,,  
\label{eq:JGUT}
\end{equation}
hence it contributes to the $\bar\theta$ angle in the mirror sector at 8$^{th}$ order
\begin{equation} \label{eq:logdivergence}
\left. \delta \bar\theta\right|_\text{mirror} \propto \frac{1}{(16\pi^2)^4}\ J_\text{GUT}\ \log \frac{\Lambda_{GUT}}{m_{\hat{H}}}\,.
\end{equation}
We can estimate the leading contribution to $J_\text{GUT}$ by considering the third and second generation only. An explicit evaluation of Eq.~\eqref{eq:JGUT} yields
\begin{equation}
    J_\text{GUT} \sim 2\ y_t^3 y_c y_b^4\ \sin^2 \theta_{23}\ \sin \alpha \sim 4\cdot 10^{-12}\ \sin \alpha\,,
\end{equation}
where $\sin \theta_{23}\sim 0.04$ is the mixing angle between these generations and $\alpha$ is one of the new CP violating phases (assumed to be potentially ${\cal O}(1)$). Combined with the loop factors, we find a harmless contribution $\delta \bar\theta \lesssim 10^{-20}$. In the SM sector one instead has to integrate out the color-triplet Higgs at $\Lambda_\text{GUT}$, leading to the usual correction  at 14$^{th}$ order~\cite{Ellis:1978hq,Khriplovich:1993pf} in $y_u,y_d$. Hence the running effects due to the SM-like Yukawa couplings $y_u,y_d$ are numerically negligible.

The other non-SM-like Yukawas, $y_1$, $y_{15}$ and $\xi$, contain many more CP phases and allow us to construct several new Jarlskog-type invariants. A careful enumeration shows that there is a unique invariant containing 6 Yukawas:
\begin{equation}
    J_6 = \text{Im}\ \text{Tr} \left(y_d^T y_d^\ast y_1 y_1^\dagger y_{15} y_{15}^\dagger \right)\,.
\end{equation}
This is expected to contribute to $\bar\theta$ at the 6$^{th}$ order, both in the SM and the mirror sectors. In both sectors the contributions start at $\Lambda_\text{GUT}$ and end at $f_a$, where the singlets and $\bf 15+\overline{15}$   receive their masses. The only mismatch between the contributions to $\bar\theta$ in the SM and the mirror sectors stem from the difference of the Higgs mass in the loops. Hence, we can estimate a misalignment of the order
\begin{equation}
    \left. \bar\theta \right|_\text{mirror} - \left. \bar\theta \right|_\text{SM} \propto \frac{1}{(16\pi^2)^3}\ J_6\ \log \frac{m_{\hat H}}{f_a}\,,
\end{equation}
for $m_{\hat H}>f_a$, and zero otherwise. We can estimate $J_6 \sim y_b^2 (y_1 y_1 y_{15} y_{15}) \sin\alpha'$, where $(y_1 y_1 y_{15} y_{15})$ is a generic combination of the Yukawa matrix elements (two generations are necessary for a non-vanishing effect) and $\alpha'$ a generic CP phase. Including the loop factors, we find $\delta \bar \theta \lesssim 10^{-8} (y_1 y_1 y_{15} y_{15})$, so that $(y_1 y_1 y_{15} y_{15}) < 10^{-2}$ suffices to satisfy  the experimental bound on the misalignment. Conversely, this effect may induce a value of the effective $\bar\theta$ in the SM that yields observable electric dipole moments in future experiments. Other Jarlskog invariants arise at 8$^{th}$ order and give contributions below the experimental limit. For completeness, we should mention that the SM sector receives additional finite contributions from threshold effects at the electroweak scale, shown to be negligible~\cite{Khriplovich:1985jr}.

We have shown that the Yukawa sector in Eq.~\eqref{eq:LYuk} allows for a good quality solution to the strong CP problem. Models that contain more Yukawa couplings, however, may introduce significant differences between the $\bar\theta$ angles in the two sectors.
Furthermore, as pointed out in ~\cite{Berezhiani:2000gh}, additional contributions to the Yukawa couplings may arise from operators at the Planck scale, such as the following dimension-5 one:
\begin{equation}
    \frac{k}{M_\text{Pl}}\ \psi_{10} \psi_{\overline{5}} H^\dagger \Phi + \text{mirror} + \text{h.c.}
\end{equation}
that gives additional contributions to the SM Yukawas once $\langle \Phi \rangle = \Lambda_\text{GUT}$, but not to the mirror sector Yukawas, where $\langle \hat  \Phi \rangle = 0$. The phase of the couplings $k$ would, therefore, induce a different Yukawa contribution to $\bar\theta$ between the SM and mirror sectors. Such operators could be forbidden by imposing a $\mathbb{Z}_2$ symmetry under which both $\Phi$ and $\hat\Phi$ are odd, hence pushing the contribution to a safer dimension-6 level~\cite{Berezhiani:2000gh}.

\vspace{0.5cm}

The desirable properties of the heavy axion solution  illustrated here are not limited to the minimal prototype $\SU(5)$ model discussed so far. As a starter, if one opens up the possibility that Yukawas couplings allowed by gauge invariance are forbidden by imposing the PQ symmetry, many more choices for the PQ fermion representations open up, as well as the case of the GUT-breaking scalar $\Phi$ in the adjoint representation of $\SU(5)$.   

\begin{table}[tb!]
    \begin{tabular}{|c|c|c|c|c|c|}
\hline
 & $\SU(5)$ & $U(1)_\text{PQ}$ & $\SU(3)_{10}$ & $\SU(3)_{5}$ & Mirror \\ \hline\hline
 $\psi_{10}^i$ & $\bf 10$ & $1$ & $3$ & $-$ & $\hat\psi_{10}$ \\
 $\psi_{\bar 5}^j$ & $\bf \bar 5$ & $-3$ & $-$ & $\bar 3$ & $\hat\psi_{\bar 5}$ \\
 $\psi_{24}$ & $\bf 24$ & $5$ & $-$ & $-$ & $\hat \psi_{24}$ \\\hline
 $\Phi$ & $\bf 24$ & $-10$ & $-$ & $-$ & $\hat \Phi$ \\
 $H$ & $\bf 5$ & $-2$ & $-$ & $-$ & $\hat H$ \\\hline
    \end{tabular}
    \caption{\label{tab:FPmodels} Matter content of a minimal mirror $\SU(5)$ GUT model realizing an invisible GUT axion, inspired by Refs~\cite{FileviezPerez:2007bcw,FileviezPerez:2019fku}.}
\end{table}

Another direction involves models where $\Phi$ carries PQ charges, hence $N f_a = \Lambda_\text{GUT}$ leading to the well-known GUT invisible axion~\cite{Wise:1981ry,Lazarides:1981kz}. In our mirror GUT setup, the axion would be generated by $\Phi$ in the visible sector, while $\hat\Phi$ acquires a large mass: as a consequence, any fermion field whose mass requires the PQ breaking would remain massless in the mirror sector. The presence of a common PQ symmetry in the two sectors can be guaranteed by introducing a portal coupling of the form $\Phi^2 (\hat\Phi^\dagger)^2$ (for $\Phi$ in a real gauge representation) or singlet fields. A minimal model can be constructed, inspired by the work in~\cite{FileviezPerez:2007bcw,FileviezPerez:2019fku}, with field content summarized in Table~\ref{tab:FPmodels}.
The Yukawa couplings consist of
\begin{multline}
    - \mathcal{L}_\text{Yuk} = \Big( y_u\ \psi_{10} \psi_{10} H + y_d\ \psi_{10} \psi_{\bar 5} H^\dagger + y_{24}\ \psi_{\bar 5} \psi_{24} H + \\
     \lambda\ \psi_{24} \psi_{24} \Phi + \mbox{mirror symmetric} \Big)\ +  \mbox{h.c.} \label{eq:LYuk2}
\end{multline}
The $\psi_{24}$ in the SM sector can be used to generate neutrino masses, while the mirror $\SU(5)$ sector below $\Lambda_\text{GUT}$ would contain a massless $\hat\psi_{24}$ in addition to the usual $\bf 10 + \bar{5}$ families. This mirror sector would also run towards strong coupling. However the low energy dynamics of this theory with an additional adjoint fermion is currently unknown, it might be confining or potentially run to an IR fixed point producing an interacting conformal field theory. 

A realization of the PQ symmetry {\`a} la DSFZ~\cite{Zhitnitsky:1980tq,Dine:1981rt} could be implemented along the lines of the Wise-Georgi-Glashow model~\cite{Wise:1981ry}, where the axion is generated by $\Phi$, as in the original proposal, or by a singlet $\phi_1$. The PQ anomalous charges are now carried by the chiral family fields $\psi_{10}$ and $\psi_{\bar 5}$ (and their mirror partners) thanks to the introduction of two Higgs multiplets with opposite PQ charges. Neutrino masses could then be included by adding singlets $\psi_1$, or an adjoint~\cite{Bajc:2006ia,DiLuzio:2018gqe}, with appropriate PQ charges.

GUTs based on larger groups, such as $\SO(10)$ and $E_6$, pose more challenges, mainly due to the fact that all SM fermions are embedded in the same representation, hence multiple Higgs representations are necessary to split the quark Yukawas and they may induce a different phase structure between the SM and mirror GUT sectors. This question can only be addressed via specific models.

\vspace{0.5cm}

A general feature of our scenario is the spontaneous breaking of the mirror $\mathbb{Z}_2$ symmetry at the GUT scale, which would lead to the cosmologically disastrous formation of domain walls~\cite{Zeldovich:1974uw,Kibble:1976sj}. To prevent the cosmological dominance of the domain wall network, they would need to decay and annihilate in the presence of a bias between the two degenerate vacua whose splitting should be at least
\begin{equation}
    \Delta V \gtrsim \frac{\sigma^2}{M_\text{Pl}^2} \sim \frac{\Lambda_\text{GUT}^6}{M_\text{Pl}^2}\,,
\end{equation}
where $\sigma \sim \Lambda_\text{GUT}^3$ is the domain wall tension. In principle, this constraint could be fulfilled by adding
dimension-5 or 6 mirror parity violating operators suppressed by the Planck scale. These additional operators would however ruin the effectiveness of the solution to the strong CP problem. Hence, the only viable possibility is for the reheat temperature after inflation to be smaller than the GUT scale, so that the domain wall network does not form. 

The breaking of the PQ symmetry also generates domain walls at the confinement scale $f_5$ of the mirror sector, depending on the discrete number $N$, c.f. Eq.~\eqref{eq:N}, that labels the degeneracy of the vacua~\cite{Sikivie:1982qv,Huang:1985tt}. For $N > 1$, like in our prototype model where $N=7$, the domain wall network is stable. The domain wall tension can be estimated to be $\sigma \sim m_a f_a^2$, typically much larger than the value that would avoid cosmological dominance ($\sigma \lesssim 10^{-5}~\text{GeV}^3$). Hence, a bias needs to be introduced, for example in the form of Planck suppressed operators already discussed in relation to the axion quality problem. Alternatively the reheat temperature could just be below the scale $f_5$. For $N=1$ the domain walls annihilate rapidly, hence no problem arises. This would be the case  for a variant of our prototype model that uses $\bf 5+\overline{5}$ as PQ fermions instead of the symmetric and these models could have a reheat temperature above $f_5$.

 \begin{figure}
     \centering \includegraphics[width=0.95\linewidth]{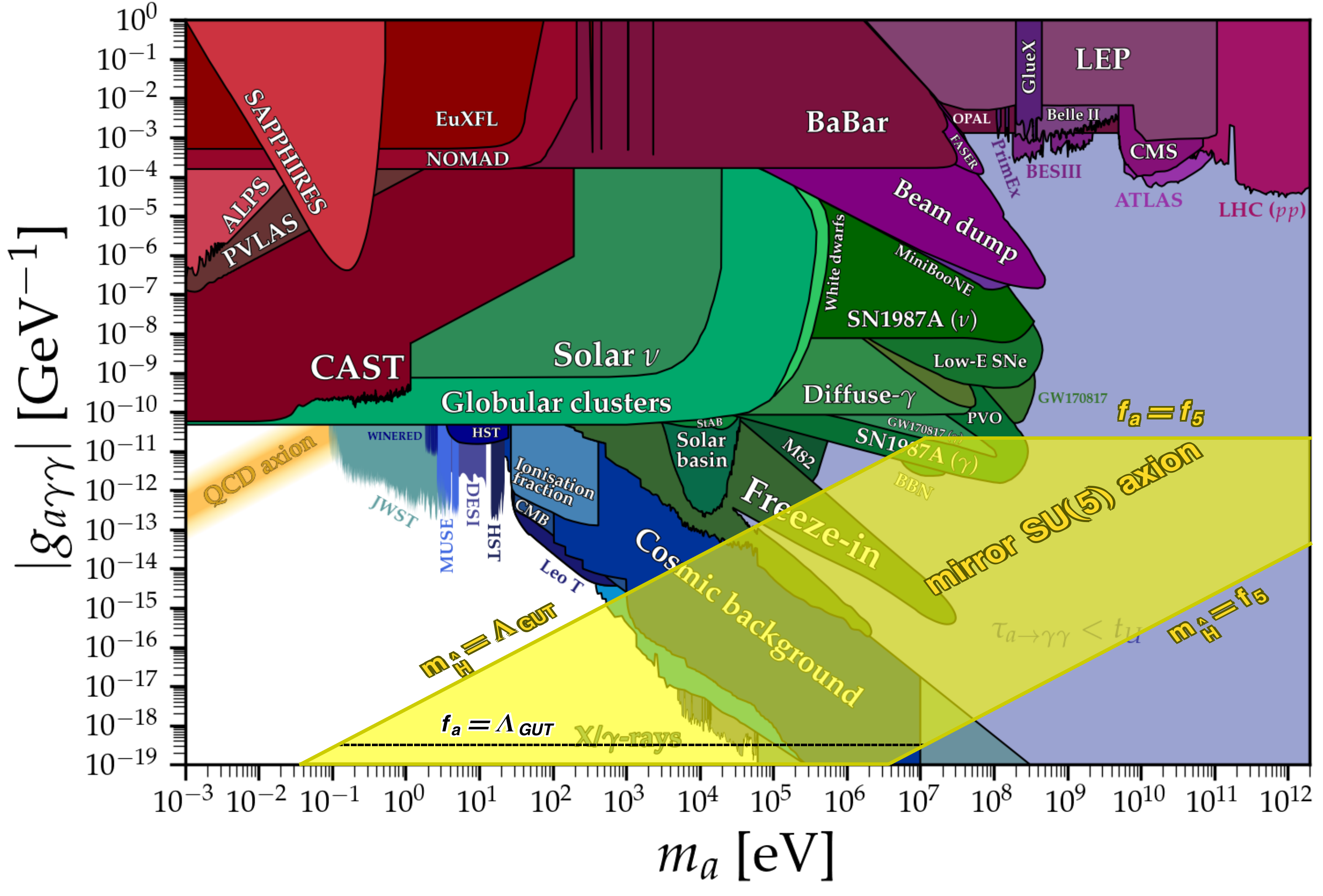}
     \caption{Parameter space of the mirror-GUT axion model in the axion--photon coupling $|g_{a\gamma\gamma}|$ versus the axion mass $m_a$, confronting experimental bounds. The yellow parallelogram shaped band shows the prediction of the mirror $\SU(5)$ axion model, with its linear oblique boundary corresponding to $m_{\hat H}=\Lambda_{\rm GUT}$ and $m_{\hat H}=f_5$, respectively. 
The horizontal upper boundary corresponds to $f_a=f_5$ and encodes the lower bound $f_a>f_5$ through Eq.~(\ref{eq:photon_coupling}). 
The lower horizontal dashed black line marks $f_a=\Lambda_{\rm GUT}$. 
Colored regions denote existing laboratory, astrophysical, and cosmological bounds, using the results and figure provided in~\cite{AxionLimits}. }
     \label{fig:axion_bounds}
 \end{figure}

\vspace{0.5cm}

Axions that address the strong CP problem are tightly constrained by observations in many different domains. The most constraining coupling is the one to photons: as in our models the anomaly is generated by complete $\SU(5)$ multiplets, the axion coupling to two photons is given by the standard formula~\cite{FileviezPerez:2019fku}
\begin{equation} \label{eq:photon_coupling}
    g_{a\gamma\gamma} = \frac{\alpha_\text{em}}{2\pi f_a}  \frac{8}{3} \sim \frac{3\cdot 10^{-3}}{f_a}\,.
\end{equation}
Note that there is no contribution due to axion-pion mixing, since the axion is predominantly getting its potential from the confining sector. This simple, model-independent, result allows us to directly compare the predicted parameter region in $m_a$ vs $f_a$ with the current exclusion limits~\cite{AxionLimits}. This is shown in Fig.~\ref{fig:axion_bounds}, where the allowed region (yellow) from Fig.~\ref{fig2} is superimposed on the bounds provided in~\cite{AxionLimits} from collider (purple), direct axion searches (red), astrophysical observations (green) and dark matter signals (blue). We see that the predicted region in our prototype model lies below those of the main exclusions, with some parts disfavored by supernova SN1987A~\cite{Hoof:2022xbe,Caputo:2022rca} and freeze-in constraints. The blue band excluded region that cuts through the center of our model region assumes that the axion saturates the observed dark matter abundance~\cite{Cadamuro:2011fd,PorrasBedmar:2024rzi,Langhoff:2022bij}. The model region to the left of the blue CMB exclusion could still provide axion dark matter. In the model region to the right at high axion masses the axion is not cosmologically stable and can't be dark matter, however it is still viable as a solution to the strong CP problem. The large value of $f_a$ will keep these heavy axion out of reach of collider probes.

\vspace{0.5cm}

In summary, we presented a new framework that allows for a heavy QCD axion, based on the idea of mirror symmetry applied to GUTs. The main novel ingredient is the unbroken mirror GUT, where a new scale higher than that of QCD is generated dynamically via the confinement of the mirror gauge symmetry. Hence, a heavy axion mass is generated without fine tuning and the new scale can be predicted given the dynamics of the SM GUT model. The large axion mass and lower decay constant also significantly relax the axion quality problem.

We proposed a simple prototype model that showcases the main features of this framework. We carefully checked that the equality of the $\bar\theta$ angles in the two sectors is not spoiled, as long as the non-SM Yukawas in the GUT model feature a mild hierarchy: this could also lead to observable electric dipole moments in the near future.  This mechanism opens the door to several interesting GUT model building directions, which we leave for future explorations. The confining mirror GUT sector generically also contains light composite states, in the form of baryons needed for the 't Hooft anomaly matching and pNGBs arising from the spontaneous breaking of the global symmetries, in addition to the axion itself. They constitute a rich hidden sector, which could host a dark matter candidate and should be studied in detail. Furthermore, the spontaneous breaking of the mirror symmetry always induces unavoidable domain walls, whose cosmological dominance must be tamed by a post-inflationary reheat temperature lower than the GUT scale. Nevertheless, it would be interesting to characterize the observable signatures of such a domain wall: as the SM interactions are confined at a high scale on the other side of the wall, the wall is expected to reflect all SM particles, acting as a {\it bona fide} cosmological mirror. Finally, this framework implies various symmetry breaking phase transitions, such as the breaking of the Peccei-Quinn symmetry and the mirror GUT confinement, which could provide observable gravitational wave imprints.

Our proposal, therefore, offers new directions in GUT model building, axion phenomenology, dark matter and cosmology, calling for further investigations.

\vspace{1cm}

\paragraph{Acknowledgements.}

\noindent
CC is supported in part by the NSF grant PHY-2309456 and also in part by the US-Israeli BSF grant No 2024091.
T.M. is partly supported by Chinese Academy of Sciences Pioneer Initiative ``Talent Introduction Plan'' (grant No.\,E4ER6601A2), the Fundamental Research Funds for the Central Universities (grant No.\,E4EQ6602X2), and the National Natural Science Foundation of China (grant No.\,E514660101).
\vspace{0.3cm}

\bibliography{biblio}

%\widetext
%\newpage

%\section*{Appendix}

%\renewcommand\thetable{S--\arabic{table}}
%\setcounter{table}{0}
%\renewcommand\thepage{\roman{page}}
%\setcounter{page}{1}

%\input{appendix}

\end{document}